\documentstyle[preprint,aps,epsf]{revtex}

\begin{document}

\draft


\title{Coupled-channels analysis of the $^{{\bf 16}}$O+$^{{\bf 208}}$Pb
fusion barrier distribution}
\author{C.R. Morton, A.C. Berriman, M. Dasgupta, D.J. Hinde, and J.O. Newton}
\address{Department of Nuclear Physics, Research School of Physical 
Sciences and Engineering,\\ Australian National University, Canberra, 
ACT 0200, Australia}
\author{K. Hagino}
\address{Institute for Nuclear Theory, Department of Physics, University of 
Washington, Seattle, WA 98195-1550}
\author{I.J. Thompson}
\address{Physics Department, University of Surrey, Guildford GU2 5XH, 
United Kingdom}
\date{Received 28 April 1999, Accepted for publication in Phys. Rev. {\bf C}}

\maketitle

\begin{abstract}
Analyses using simplified coupled-channels models have been unable to 
describe the shape of the previously measured fusion barrier 
distribution for the doubly magic $^{16}$O+$^{208}$Pb system.
This problem was investigated by re-measuring the fission excitation 
function for $^{16}$O+$^{208}$Pb with improved accuracy and performing 
more exact coupled-channels calculations, avoiding the constant-coupling 
and first-order coupling approximations often used in simplified analyses. 
Couplings to the single- and 2-phonon states of $^{208}$Pb, correctly taking 
into account the excitation energy and the phonon character of these states, 
particle transfers, and the effects of varying the diffuseness of the nuclear 
potential, were all explored.  However, in contrast to other recent analyses 
of precise fusion data, no satisfactory simultaneous description of the shape 
of the experimental barrier distribution and the fusion cross-sections for  
$^{16}$O+$^{208}$Pb was obtained.
\end{abstract}

\pacs{PACS number: 25.70.Jj, 24.10.Eq, 21.60.Ev, 27.80.+w}

\section{Introduction}
\label{intro}

Precise fusion cross-sections have been measured for many reactions, 
involving nuclei which exhibit different collective degrees of freedom.
Their excitations, through coupling to the relative motion of the colliding 
nuclei, cause a splitting in energy of the single fusion barrier resulting 
in a distribution of barriers, which drastically alters the fusion 
probability from its value calculated assuming quantal tunnelling through 
a single barrier.  It was shown by Rowley {\em et al.}~\cite{Rowley91} that, 
under certain approximations, the distribution in energy of a discrete 
spectrum of barriers could be obtained from precise fusion cross-sections 
$\sigma$ by taking the second derivative with respect to the center-of-mass 
energy $E_{\mbox{\scriptsize c.m.}}$ of the quantity 
$(E_{\mbox{\scriptsize c.m.}} \sigma)$.  When the effects of quantal 
tunnelling are considered, 
$d^2(E_{\mbox{\scriptsize c.m.}}\sigma)/d E^2_{\mbox{\scriptsize c.m.}}$
becomes continuous, and each barrier is smoothed in energy with a full width 
at half maximum (FWHM) of $0.56 \hbar \omega$, where $\hbar \omega$ is the 
barrier curvature.  
The difference between a more realistic calculation of 
$d^2(E_{\mbox{\scriptsize c.m.}}\sigma)/d E^2_{\mbox{\scriptsize c.m.}}$ 
(where the angular momentum dependence of the curvature and barrier radius 
is taken into account) and the smoothed barrier distribution is 
small~\cite{Leigh95}, and so it is convenient to refer to
$d^2(E_{\mbox{\scriptsize c.m.}}\sigma)/d E^2_{\mbox{\scriptsize c.m.}}$
as the fusion barrier distribution.

The fusion barrier distribution can be a very sensitive `gauge' of the 
dominant collective modes excited during the collision~\cite{Leigh95}.  
Its shape is related to the nuclear structure of the reactants.  Barrier 
distributions have been measured for nuclei with static deformations 
\cite{Leigh95,Wei91,Lemmon93,Leigh93,Hinde95,Bierman96,Bierman96b,Mein97},
for nuclei where vibrational degrees of freedom 
dominate~\cite{Leigh95,Morton94}, in systems where the effects of transfer 
channels~\cite{Leigh95,Morton94,Timmers97} and multi-phonon 
excitations~\cite{Stefanini95,Hinde95b} are important, and where the 
influence of the projectile excitation is 
prominent~\cite{Bierman96,Bierman96b,Hinde95b}.  

The precise fusion data have stimulated advances in the quantitative 
application of the coupled-channels (CC) description of fusion, and many 
experimental barrier distributions have been well reproduced with various 
degrees of refinement of this model.
The CC description is expected to be simpler for systems involving the 
fusion of closed-shell nuclei due to the presence of relatively few 
low-lying collective states. 
An example is the $^{16}$O+$^{144}$Sm system, where a good 
description~\cite{Morton94} of the experimental barrier distribution was 
obtained with a simplified CC model~\cite{ccfus,Dasgupta92}.  
This description was somewhat fortuitous in view of the approximations 
used in this model.  An improvement in the description of the barrier 
distribution was achieved with more exact CC 
calculations~\cite{Hagino97,Hagino97b,Hagino97c} which correctly treated 
the excitation energies and the phonon character of the coupled states. 

Given the current level of knowledge of the theoretical description of 
heavy-ion fusion, and the success of calculations in reproducing 
the shape of the measured barrier distribution for $^{16}$O on 
$^{144}$Sm, it might be expected that present models should be able to 
describe the fusion of $^{16}$O with the doubly-magic nucleus $^{208}$Pb.
The $^{16}$O + $^{208}$Pb system is also one of the few cases where there 
is existing knowledge of important particle transfer channels.
The fusion barrier distribution for the $^{16}$O + $^{208}$Pb reaction has 
been measured previously~\cite{Morton95}, however it was not possible to 
obtain an adequate theoretical description of its shape.  This could have 
been due to shortcomings in the experiment or the simplified CC analysis 
used in calculating the theoretical barrier distribution.  Improvements 
in the available techniques of precise fission 
cross-section measurements, including the use of fragment-fragment 
coincidences, were reason to re-measure the fusion excitation function for 
the $^{16}$O + $^{208}$Pb reaction. 

The purpose of the current work was to find the cause of the previous 
disagreement between theory and data by comparing the newly measured 
barrier distribution with more exact CC calculations, and to identify the 
dominant couplings in the fusion of $^{16}$O+$^{208}$Pb.
The coupled-channels analysis of the new fusion data has 
proved to be more difficult than expected, and a complete description of 
the data has not yet been obtained.

\section{Experimental Results}
\label{exper}

The re-measurement of the fission excitation function for 
$^{16}$O + $^{208}$Pb was performed at the Australian National University 
using $^{16}$O beams from the 14UD Pelletron accelerator.  The beams were 
pulsed with bursts of $1$ ns FWHM, separated by 
$106.6$ ns.  Beam energies used were in the range $75$--$118$ MeV, in 
increments of $0.6$ MeV up to $88$ MeV.  
The absolute beam energy was defined to better than $0.05$ 
MeV and the relative beam energy to better than a few keV~\cite{Leigh95}.  
The target was $40$--$45$~$\mu$gcm$^{-2}$ of $^{208}$PbO deposited on a 
backing of $\approx 10~\mu$gcm$^{-2}$ of C.  The isotopic purity of the 
$^{208}$Pb was $99.0 \pm 0.1 \%$. 
Fission fragments were detected in two of the large-area 
multiwire proportional counters (MWPCs) of the CUBE detector system.  
One was positioned in the backward hemisphere covering the scattering angles 
$-171^{\circ} \leq \theta_{\mbox{\scriptsize lab}} \leq -94^{\circ}$, and 
the other in the forward hemisphere with $4^{\circ} \leq 
\theta_{\mbox{\scriptsize lab}} \leq 81^{\circ}$.  The fission fragments 
were identified in an individual detector by their energy loss signal, 
and the time-of-flight measured relative to the pulsed beam.

In the measurement described in Ref.\ \onlinecite{Morton95}, only a single 
MWPC located in the backward hemisphere was used.  However, in the present 
measurement, the front MWPC was operated in coincidence with the back 
fission detector, and the fission fragments were identified with the 
time-of-flight in one detector versus the time-of-flight in the other.
This allowed good separation between the fission events from the 
$^{16}$O+$^{208}$Pb reaction and other reactions with the target, which 
were a problem for the low cross-sections in the earlier measurement. 
The fission cross-section was measured down to energies where the 
evaporation residue cross-sections were previously 
determined~\cite{Morton95}.  Two silicon surface--barrier detectors, 
located at $\pm 22.5^{\circ}$ to the beam axis, were used to monitor 
the Rutherford scattering for normalisation of the fission fragment yield. 
The fission fragment yields in the MWPCs were converted into 
fission cross-sections as described in 
Refs.~\cite{Morton95,Hinde95b}.

The new fission excitation function is shown in Fig.~\ref{fig1}(a), 
together with the results from the previous 
measurement~\cite{Morton95}, as indicated by the open circles in 
Fig.~\ref{fig1}(a).  The fusion cross-sections $\sigma$ for 
$^{16}$O + $^{208}$Pb were obtained by summing 
$\sigma_{\mbox{\scriptsize fis}}$ and the 
evaporation residue cross-sections published in Ref.\ \onlinecite{Morton95}, 
interpolating where necessary.  The present data (solid circles) and 
previously published fusion cross-sections (open circles) are shown in 
Fig.~\ref{fig1}(b).  The fusion cross-sections from the new measurement are 
presented in Table~\ref{Table1}.

The fusion barrier distribution was obtained by evaluating the point 
difference formula of Ref.\ \onlinecite{Leigh95} using an energy
step of $\Delta E_{\mbox{\scriptsize c.m.}}=1.67$ MeV.  The resulting 
barrier distribution is shown in Fig.~\ref{fig2} by the solid circles.  For 
comparison, the barrier distribution (open points) in 
Ref.\ \onlinecite{Morton95} is reproduced, where each symbol represents one 
of the three separate passes through the fusion excitation function.  In 
Ref.\ \onlinecite{Morton95}, the barrier distribution was calculated with 
$\Delta E_{\mbox{\scriptsize c.m.}}=1.86$ MeV.  The difference in the two 
step lengths does not have any 
significant effect on the calculated barrier distributions since they are 
already smoothed by $\approx 2$ MeV due to quantum tunnelling 
effects~\cite{Rowley91}.

The new data are generally in good agreement with the previous measurement, 
but give a better defined barrier distribution.  
This is mainly due to the improved statistics, the clean identification of 
fission events made possible by operating two detectors in coincidence, 
and better definition and consistency of the angle between the beam axis and 
the fission detectors.
The slight disagreement between the two barrier distributions 
can be largely attributed to three errant points in the original excitation 
function at $E_{\mbox{\scriptsize c.m.}}=73.8$, $74.3$ and $75.2$ MeV, which 
differ from the current data by up to $5\%$.
Since $d^2(E^i_{\mbox{\scriptsize c.m.}}\sigma)/d 
E^{i2}_{\mbox{\scriptsize c.m.}}$ at an energy 
$E^i_{\mbox{\scriptsize c.m.}}$ is evaluated with a three--point 
difference formula, each wayward cross-section affects a total of three 
points, that point at $E^i_{\mbox{\scriptsize c.m.}}$, and its two 
neighbouring points at $(E^i_{\mbox{\scriptsize c.m.}} \pm 1.67)$ MeV.  
For example, the old cross-section at $E_{\mbox{\scriptsize c.m.}}=75.2$ 
MeV was high with respect to the new measurement.  This means that 
$d^2(E_{\mbox{\scriptsize c.m.}}\sigma)/d E^2_{\mbox{\scriptsize c.m.}}$ at 
$E_{\mbox{\scriptsize c.m.}}=75.2$ MeV is lower than the new 
barrier distribution, and $d^2(E_{\mbox{\scriptsize c.m.}}\sigma)/d 
E^2_{\mbox{\scriptsize c.m.}}$ at both $E_{\mbox{\scriptsize c.m.}}=73.3$ 
MeV and $77.1$ MeV are high (see the encircled points in Fig.~\ref{fig2}).  

Nevertheless, the general features, such as the height of the main peak, 
and shape of the two barrier distributions are in good agreement.   

\section{Coupled-channels analysis of the measured fusion barrier 
distribution}
\label{cc}

Several ingredients are required for a coupled-channels description of the
fusion barrier distribution.  Inputs to the model calculations include the 
nucleus-nucleus potential parameters, the coupling strengths of the 
vibrational states and their excitation energies.  In addition, there are 
choices to be made regarding various assumptions and approximations used in 
the solution of the coupled equations.  

\subsection{The coupled-channels calculations}
\subsubsection{Nuclear potential parameters}
\label{nucpot}

The nuclear potential parameters were determined with consideration of two 
constraints:  (i) fitting the high-energy fusion cross-sections, (ii) 
choosing a sufficiently deep nuclear potential, which is consistent 
with the ingoing-wave boundary condition used in the CC calculations.
The measured fusion cross-sections at energies above the average barrier 
were fitted using a single--barrier penetration model, with an 
energy-independent nuclear potential, Woods-Saxon in form, with 
\begin{equation}
  V(r)=-V_0/(1+\exp[(r-r_0 A_P^{1/3}-r_0 A_T^{1/3} )/a]),
\end{equation}
where $V_0$ is the depth, $r_0$ is the radius parameter, and $a$ is the
diffuseness of the nuclear potential.  With $V_0$ chosen to be $50$ MeV, 
$r_0$ and $a$ were varied to obtain the best fit to $\sigma$.  This resulted 
in the parameters $V_0=50$ MeV, $r_0=1.159$ fm and $a=1.005$ fm, giving
an average barrier $B_0=74.5$ MeV at a barrier radius of $R_B=11.3$ fm with 
curvature for the average barrier of $\hbar \omega_0 =3.07$ MeV.
The excitation function and fusion barrier distribution associated with 
these single--barrier (SB) parameters are shown by the dot-dot-dashed 
lines in Fig.~\ref{fig3}(a) and (b), respectively.  

The above values for $V_0$ and $r_0$ could not be used in the CC codes 
because the potential depth was too shallow causing high-$\ell$ partial 
waves that should have been absorbed (contributing to the 
fusion cross-section) to be reflected at the barrier. 
To ensure that all the ingoing flux was absorbed inside the fusion 
barrier, a new set of potential parameters was obtained with the 
diffuseness parameter fixed at $a=1.005$ fm, and $V_0$ was increased to 
$200$ MeV, compensated by a reduction in $r_0$ to $0.978$ fm to obtain 
the same fusion barrier $B_0=74.5$ MeV, which occurs at $R_B=11.5$ fm 
with a curvature $\hbar \omega_0 =3.87$ MeV.  By making this adjustment 
in $V_0$, the quality of the fit to the high-energy fusion 
cross-sections is reduced.  However, this is not of concern for the 
following reasons.

The main aim of this analysis is the reproduction of 
the {\em shape} of the measured barrier distribution, a quantity which is
insensitive to small changes in the potential parameters.   In comparison, 
the high-energy fusion cross-sections are very sensitive to height of the 
average barrier, and can always be fitted by adjusting the potential 
parameters.  However, since there exists some 
sensitivity of the calculated high-energy fusion cross-sections to the 
couplings~\cite{Leigh95}, this would mean the nuclear potential parameters 
would need to be adjusted for each different coupling 
scheme if the fit to the high energy data is to be retained.  
Rather than re-fitting the high energy data after each new coupling scheme, 
the CC calculations were performed without any further adjustment to the 
bare nuclear potential.  This meant that the calculated fusion 
cross-sections overestimated the data in the high energy region, see 
for example the CC calculations in Fig.~\ref{fig3}(a).  The data in the 
high energy region could be re-fitted with a slightly higher average 
fusion barrier, corresponding to a different set of potential parameters, 
but this would cause only a shift up in energy of the whole barrier 
distribution, without any appreciable change in its shape. 

The diffuseness parameter obtained from the above procedure is 
significantly larger than that deduced from elastic scattering 
measurements~\cite{Christ76}, a result common to other fusion 
analyses~\cite{Leigh95}.
The inconsistency between the diffuseness parameters obtained from fusion 
and elastic scattering data implies that the potential parameters obtained 
are specific to the data being fitted.
It is also possible that the potential parameters obtained from a fit to 
the data in the high energy region are not applicable 
at energies in the barrier region, or below the lowest barrier.  
In this sense the potential parameters obtained are {\em effective} ones, 
and the true interaction potential remains an uncertainty in these 
calculations.

The effect of using a smaller diffuseness is shown in 
Fig.~\ref{fig3}(a) and (b), where two calculations are compared, one with 
$a=0.65$ fm and the other with 
$a=1.005$ fm, both with the same average barrier $B_0=74.5$ MeV.  Couplings 
to the single phonon states in $^{208}$Pb are included in these CC 
calculations (see Sec.~\ref{single}).  
For $E_{\mbox{\scriptsize c.m.}}<B_0$, the cross-section 
for the calculation with $a=0.65$ fm falls less rapidly than the $a=1.005$ 
fm case, since the smaller diffuseness gives a narrower barrier (larger 
$\hbar \omega_0$) and hence a larger barrier penetrability.  
In the barrier region, a smaller diffuseness reduces the height of the 
main peak in
$d^2(E_{\mbox{\scriptsize c.m.}}\sigma)/d E^2_{\mbox{\scriptsize c.m.}}$, 
due to the increase in the width of the tunnelling factor~\cite{Rowley91} 
which smooths the barrier distribution (see Sec.~\ref{smalla}).  
These calculations demonstrate the effect on the calculated barrier 
distribution of the uncertainty in the appropriate choice of the 
diffuseness parameter.  Further experiments are required to address 
this problem.

\subsubsection{Approximations used in solving the coupled equations}

In the coupled-channels calculations that follow, except for the 
{\tt FRESCO} calculations, the no-Coriolis or isocentrifugal 
approximation~\cite{Lindsay84,Tanimura85,Kruppa93,Hagino95} was used.  
This approximation has been 
shown~\cite{Esbensen96,Tanimura87} to be good for heavy-ion fusion 
reactions.  The calculations included coupling to all orders in the 
deformation parameter for the nuclear coupling matrix.  
In the past, when making quantitative comparisons with the fusion data, 
the linear coupling approximation was often used.  Here the nuclear 
coupling potential was expanded with respect to the deformation parameter 
keeping only the linear term.  
It was shown~\cite{Hagino97,Esbensen96,Esbensen87} that the 
agreement between the measured and calculated fusion cross-sections was 
improved with the inclusion of second-order terms.
Later, Hagino {\em et al.}~\cite{Hagino97} demonstrated that, for 
heavy symmetric systems at least, the effect of inclusion of terms higher 
than second-order in the nuclear coupling potential was as significant as 
including the second-order term itself.
Even though this effect was largest for heavy near-symmetric systems, it 
was also found to be significant for reactions involving lighter nuclei 
such as $^{16}$O+$^{144}$Sm.  

The linear coupling approximation was retained for the Coulomb coupling 
potential since inclusion of terms of higher order has been shown to have 
only a very minor effect on the barrier distribution~\cite{Hagino97}.

The excitation energies of the vibrational states were treated exactly in 
these calculations.  Consequently, there were no approximations associated 
with the eigenchannel approach used in simplified CC analyses, such as 
those present in the code {\tt CCFUS}~\cite{ccfus}.

\subsection{Channel couplings}
\subsubsection{Coupling to single-phonon states in \protect$^{208}$Pb}
\label{single}

Both $^{144}$Sm and $^{208}$Pb are spherical, vibrational nuclei with 
similar low-lying collective states, so it might be expected that the 
coupling scheme which was successful in the description of the barrier 
distribution for $^{16}$O + $^{144}$Sm would also provide a good 
description of the $^{16}$O + $^{208}$Pb reaction.
The measured barrier distribution for the $^{16}$O + $^{144}$Sm reaction 
was well described by coupling to the single-phonon states 
in $^{144}$Sm~\cite{Morton94}, where the dominant channel is the 
single-octupole phonon state.
The analogous calculation for $^{16}$O + $^{208}$Pb is shown by the solid 
lines in Fig.~\ref{fig4}(a) and (b).  The calculation was performed with 
the CC code {\tt CCFULL}~\cite{Hagino97,Hagino99}, where 
fusion is simulated using the ingoing-wave boundary condition. 
Coupling to the $3^{-}_1$ and $5^{-}_1$ single-phonon states in $^{208}$Pb 
was included, with the relevant parameters summarized in Table~\ref{Table2}.  
This calculation fails to reproduce the shape of the measured barrier 
distribution [see Fig.~\ref{fig4}(b)].  Although the calculation produces a 
two-peaked structure, mainly due to coupling to the $3^{-}_1$ state in 
$^{208}$Pb, there is still too much strength in the main peak of the 
theoretical barrier distribution, which implies that more coupling is 
required.

Additional coupling to other single-phonon states in $^{208}$Pb produced no 
improvement in the agreement with the measured barrier distribution, due to 
the relative weakness of these couplings.  
In relation to the disagreement between theory and the data in 
Fig.~\ref{fig4}(b), an initial impression is that the area of the calculated 
barrier distribution is larger than that of the measurement.  This difference 
could be caused by a lower fusion yield resulting from a loss of flux due to 
incomplete fusion.  Such an effect was recently observed~\cite{Dasgupta99} in 
the fusion of $^{9}$Be on $^{208}$Pb.  However, evaluation of the area under  
$d^2(E_{\mbox{\scriptsize c.m.}}\sigma)/d E^2_{\mbox{\scriptsize c.m.}}$, 
a quantity which should be approximately proportional to the geometric area 
$\pi R_B^2$, indicates that this is not the case.
The area under the theoretical barrier distribution represented by the solid 
line in Fig.~\ref{fig4}(b) is $4227$ mb, implying a value of 
$R_B=11.6$ fm for the average barrier radius, obtained by simply equating 
the area with $\pi R_B^2$.   This compares with the area under the 
experimental barrier distribution of $3981$ mb, implying a radius 
$R_B=11.3$ fm.  The difference between the theoretical and experimental 
areas is only $6 \%$, $3.6 \%$ of which is due to use of the larger 
potential depth, $V_0=200$ MeV, which has a radius $R_B=11.5$ fm instead 
of the best fit value of $R_B=11.3$ fm for $V_0=50$ MeV.  Thus, the 
mismatch between experiment and theory to the level of $\approx 2$--$3\%$, 
is not due to incomplete fusion.

To obtain a successful theoretical description of the $^{16}$O+$^{208}$Pb 
reaction, a coupling scheme that produces a barrier distribution with a 
shape corresponding to the measured one is required.  
Since the areas under the experimental and theoretical barrier 
distributions are in good agreement, the height of the 
main barrier in the distribution will be used as an indicator of the 
ability of theory to reproduce the overall shape of the experimental 
barrier distribution.

\subsubsection{The effects of coupling to particle transfers}
\label{transfer}

Attempts have been made previously to `explain' qualitatively deviations 
between theory and experiment as being due to neglect of transfer couplings.  
Such an approach has been taken because of the difficulty of treating the 
transfer
process in a realistic way, and the lack of knowledge of transfer coupling 
strengths.  However in $^{16}$O + $^{208}$Pb, some of the important transfer 
coupling strengths have been measured.  To ascertain the significance of the 
effects of transfer couplings on fusion, both the transfer and inelastic 
channels (with coupling to all orders) should be considered simultaneously 
in the CC calculation.  
The effects of particle transfers on the fusion cross-sections and spin 
distributions for $^{16}$O + $^{208}$Pb have been calculated by Thompson 
{\em et al.}~\cite{Thompson89} at 8 energies between $E_{\mbox{\scriptsize 
lab}}=78$ and $102$ MeV, using the coupled-channels code 
{\tt FRESCO}~\cite{Thompson88}.  
Here, those calculations have been repeated, 
with a minor modification to the nuclear potential in the entrance-channel 
mass partition, with coupling to all orders in the nuclear potential, and 
with smaller energy steps in order to obtain the barrier 
distribution.  This was necessary since it was not possible to treat 
transfer correctly using the code {\tt CCFULL}.  The details of this 
calculation are discussed below.

Before proceeding with the transfer calculations, the results of the two 
coupled-channels codes used in this work were compared.   
The comparison was made with a {\tt FRESCO} calculation using parameters 
identical to the single-phonon 
calculation described in Sec.~\ref{single}.  The {\tt FRESCO} 
calculation was performed with version {\tt FRXX}, which includes a new 
option allowing coupling to all orders in the nuclear coupling potential, 
as in the calculation described in Sec.~\ref{single}.
The barrier distribution from {\tt FRESCO} is shown by the dashed line 
in Fig.~\ref{fig4}(b).  There is very good agreement between it and the 
barrier distribution calculated using {\tt CCFULL} [solid 
line in Fig.~\ref{fig4}(b)].  The small difference between the solid and 
dashed lines in Fig.~\ref{fig4}(b) may be due to the isocentrifugal 
approximation which was used in the {\tt CCFULL} calculation.  

Having established the agreement between the above two calculations for 
inelastic couplings, the effects of coupling to transfer channels were 
examined with {\tt FRESCO}.  In addition to the inelastic couplings, the 
following three transfer couplings were included, which are those included 
in the previous analysis~\cite{Thompson89}.
The single-neutron pickup reaction 
($^{16}$O,$^{17}$O) with $Q=-3.2$ MeV, the single-proton stripping reaction 
($^{16}$O,$^{15}$N) with $Q=-8.3$ MeV and the 
$\alpha$-stripping reaction ($^{16}$O,$^{12}$C), where $Q=-20$ MeV, were 
included.  The spectroscopic factors for the single-nucleon transfers were 
taken from Ref.\ \onlinecite{Franey79}, and in the case of the 
$\alpha$-stripping 
couplings, were set to reproduce the measured transfer yield.  Coupling to 
excited states in $^{17}$O, $^{15}$N, $^{207}$Pb and $^{209}$Bi was 
included as described in Ref.\ \onlinecite{Thompson89}. The real and 
imaginary potential parameters for all three transfer partitions were 
$V_0=78.28$ MeV, $r_0=1.215$ fm, $a=0.65$ fm and $V_i=10$ MeV, 
$r_{0i}=1.00$ fm, $a_i=0.40$ fm, respectively.

The barrier distribution from the {\tt FRESCO} calculation including 
transfer is shown by the dot-dot-dashed line in Fig.~\ref{fig4}(b).  
Compared to the case with no transfer, the main peak of the barrier 
distribution is shifted down in energy and its height is reduced, whilst 
the second peak in the distribution is smoothed in energy.
Of the three transfer couplings considered in this calculation, the 
neutron--pickup transfer has the largest effect on the barrier distribution, 
since it is the most strongly populated transfer.  
Using a set of potential parameters for the 
$^{17}$O+$^{207}$Pb mass partition different to those quoted above, with a 
real diffuseness of $a=1.005$ fm, had only a small effect on the shape of 
the barrier distribution. 
The $0.5$ MeV shift downwards in energy of the barrier distribution is not 
problematic, since there is freedom to renormalise the bare potential to a 
value which will shift the theoretical barrier distribution back to its 
original position.
Of importance here is the ability to reproduce the shape of the barrier 
distribution, and although the coupling to the transfer channels reduces 
the height of the main peak in the barrier distribution, it is not 
sufficient, implying that further couplings are required.  

Additional transfer channels, which have been neglected in the present 
calculation, are unlikely to significantly improve the agreement, since the 
above three transfer couplings represent the most strongly populated 
transfers.  The effects of additional transfers on the fusion cross-section 
were investigated in Ref.\ \onlinecite{Sastry98}, where it was found that 
the $\alpha$- and triton-pickup transfers had no effect on $\sigma$.  The 
2-neutron pickup, with $Q=-1.9$ MeV, did affect the fusion cross-section, 
although the increase in $\sigma$ was at most a factor of $1.11$ above the 
calculation without this transfer, at $E_{\mbox{\scriptsize lab}}=78$ MeV.  
This compares with an enhancement in $\sigma$ at the same energy of 
$\approx 2.5$ between the transfer calculation with neutron-pickup, proton 
and $\alpha$-stripping over the calculation without these transfer 
couplings.

\subsubsection{The effects of coupling to the \protect$3^-_1$ in 
\protect$^{16}$O }
\label{projectile}

The treatment of projectile excitations in CC analyses deserves some 
comment.  The measured barrier distributions for the reaction $^{16}$O 
with various isotopes of samarium~\cite{Leigh95} 
showed no specific features associated with excitation of the octupole
state in $^{16}$O.  It was shown in Ref.\ \onlinecite{Leigh95} that coupling
to the $3^-_1$ state in $^{16}$O at $6.13$ MeV using the simplified CC code
{\tt CCMOD}~\cite{Dasgupta92}, which uses the linear coupling approximation, 
resulted in a deterioration in the agreement with the measured barrier 
distribution.  This effect is related to the neglect of the higher-order 
terms in the CC calculations~\cite{Hagino97b,Esbensen96}.  Since the 
transition strength of the $3^-_1$ state in $^{16}$O is large, higher-order 
terms should be included in the expression for the nuclear 
coupling potential.  
When the $3^-_1$ state in $^{16}$O was included with coupling to all orders 
in the nuclear potential, the theoretical barrier distribution was 
essentially restored to its shape before the inclusion of the projectile 
coupling~\cite{Hagino97b}.  
However, the whole barrier distribution was shifted down in energy by a 
few MeV.  This shift has been explained~\cite{Esbensen83,Tak94,Hagino97d} 
in terms of the adiabaticity of the projectile excitation.  
When the excitation energy of a state is large, then the timescale of the 
intrinsic motion is short compared to the tunnelling time, allowing the 
projectile to respond to the nuclear force in such a way as to always be 
in the lowest energy configuration.  This means that coupling to states 
like the $3^-_1$ state in $^{16}$O, only leads to a shift in the average 
fusion barrier, and so is equivalent to a renormalisation of the effective 
potential.

In order to confirm the above result for the $^{16}$O+$^{208}$Pb reaction, 
calculations were performed with coupling to the $3^-_1$ state in $^{16}$O 
at $6.13$ MeV using the code {\tt CCFULL}.  
No better agreement with the shape of the measured 
barrier distribution resulted, causing only a shift in energy of the whole 
barrier distribution, without an appreciable change in its overall shape.
An example of this effect is shown in Fig.~\ref{fig7}(b).

In summary, the calculations described above, with a single-phonon plus 
transfer coupling scheme, were unable to describe the measured barrier 
distribution.  In the next Section, the effects of a larger coupling space 
are explored.  The following calculations result mostly from the code 
{\tt CCFULL}.  Due to the long computational time involved, {\tt FRESCO} 
was used only to estimate the additional effects of coupling to transfer 
channels.

\subsubsection{Coupling to the 2-phonon states in 
\protect$^{208}$Pb }
\label{double}

In the doubly magic nucleus $^{208}$Pb, the energy of the first $3^-$ state 
is at $2.614$ MeV and is interpreted~\cite{Hamamoto74} as a collective 
octupole state because of its large $B(E3)$ value. 
In the harmonic vibrational model, the 2-phonon state would be 
expected~\cite{BM75} at an energy twice that of the single-phonon excitation.
Hence in $^{208}$Pb, the 2-phonon state 
$[3^-_1 \otimes 3^-_1]$, consisting of the $0^+$, $2^+$, $4^+$, and $6^+$ 
quadruplet of states, is expected~\cite{Hamamoto74} at the unperturbed 
energy of $5.228$ MeV.  There have been a number of searches for members 
of the 2-phonon quadruplet, including a recent $(n,n^{\prime}\gamma)$ 
measurement~\cite{Yeh96} which found evidence for the existence of the 
$0^+$ state at $5.241$ MeV.  
A more recent measurement~\cite{Vetter98} using Coulomb excitation, did not 
identify any new state around $5.2$ MeV, but was able to extract the 
$B(E3,3_1^-\rightarrow6_1^+)$ value for the lowest known $6^+$ state at 
$4.424$ MeV, whose strength suggested a strong fragmentation of the 2-phonon 
state in $^{208}$Pb.

Because of the expected strong collective nature of the low-lying octupole 
state in $^{208}$Pb, it is likely that 2-phonon excitations play some role 
in the fusion of $^{16}$O on $^{208}$Pb. 
The effects of the inclusion of 2-phonon excitations on the fusion barrier 
distribution have been investigated theoretically by Kruppa 
{\em et al.}~\cite{Kruppa93} as well as Hagino {\em et al.}~\cite{Hagino98}.  
Recent experimental evidence has come from a
measurement of the barrier distribution for the $^{58}$Ni+$^{60}$Ni 
reaction~\cite{Stefanini95}, where it was demonstrated that fusion is 
sensitive to such complex multi-phonon excitations.

The barrier distribution shown by the solid line in Fig.~\ref{fig5} is a 
{\tt CCFULL} calculation which includes, in addition to the 
$3^-_1$ and $5^-_1$ single-phonon states in $^{208}$Pb, coupling to the 
double-octupole phonon in the target.  This calculation was performed in 
the harmonic limit, where the energy of the $[3^-_1 \otimes 3^-_1]$ state 
was taken to be $5.23$ MeV, with the strength of coupling between the 
single- and 2-phonon states given by $\sqrt{2}\beta_{3}$, the coupling 
expected in the harmonic limit.  The 2-phonon result produces a 
shoulder in the barrier distribution at 
$E_{\mbox{\scriptsize c.m.}} \approx 76$ MeV 
whilst reducing the height of the main barrier, leading to a minor 
improvement over the single-phonon coupling scheme.
The inclusion of multiple excitations in the target, for example the 
$[5^-_1 \otimes [3^-_1 \otimes 3^-_1]]$ state, did not result in any 
significant difference to the barrier distribution given by solid line 
in Fig.~\ref{fig5}, largely due to the fact that $\beta_5$ is very small.
The additional inclusion of the $3^-_1$ state in the projectile, and 
mutual excitations of the projectile and target, was also found to have 
little effect on the shape of the calculated barrier distribution.

The next obvious choice to consider is coupling to the 2-phonon states in 
$^{208}$Pb plus the transfer channels.  Such a CC calculation was performed 
with {\tt FRESCO}, and the results are shown by the dashed line in 
Fig.~\ref{fig5}.  This causes a small shift in the barrier distribution to 
lower energies and an enhancement in the height of the shoulder at 
$E_{\mbox{\scriptsize c.m.}} \approx 76$ MeV over the single-phonon plus 
transfer calculation.  Although the effect of these 
couplings are helpful, the resultant barrier distribution is still well 
short of a complete description of the data. 
One effect still not accounted for is multi-step transfer couplings.  
With the present CC codes, it was not 
possible to include transfer {\em from} the excited states in $^{208}$Pb, 
and the effect of neglecting these channels on the barrier distribution 
is not known.
However, it was possible to check if the anharmonicity of the 2-phonon 
states was responsible for the remaining disagreement.
Below, the size of these effects are estimated.

\subsubsection{The anharmonicity of the 2-phonon quadruplet in 
\protect$^{208}$\mbox{Pb}}
\label{anharmon}

When 2-phonon states were included in the coupling scheme for 
$^{16}$O+$^{144}$Sm, using the harmonic vibrational 
model, the good agreement between the measured and calculated 
barrier distribution was lost~\cite{Morton95b}.  
At first, this result was puzzling in that there is both theoretical and 
experimental evidence for the 
presence of double-octupole phonon states in $^{144}$Sm~\cite{Gatenby90}.
However, deviations from the pure harmonic vibration model are expected to 
occur and the assumption of vibrational harmonicity for the coupling in 
$^{144}$Sm is not correct.  Subsequently it was 
demonstrated~\cite{Hagino97c} within the framework of the interacting 
boson model, that when the 
anharmonicities of the double-phonon states were accounted for, the 
theoretical barrier distribution was restored to a shape matching the
experiment.  In fact, anharmonic coupling to the additional 2-phonon 
states marginally improved the agreement relative to the single-phonon  
description of the data.

It has been known for a long 
time that the $3^-_1$ state in $^{208}$Pb has a large quadrupole 
moment, which is indicative of the anharmonic effects in octupole 
vibrations~\cite{Hamamoto74}.  The anharmonic effects give rise to a 
splitting in energy of the $0^+$, $2^+$, $4^+$, and $6^+$ members of the 
2-phonon quadruplet in $^{208}$Pb.
In the Coulomb excitation search for 2-phonon states in 
$^{208}$Pb by Vetter {\em et al.}~\cite{Vetter98}, the authors found that 
the lowest lying $6^+$ state populated had a transition strength only 
$\approx 20 \%$ of the harmonic $B(E3)$ value, indicating a possible 
fragmentation of the octupole vibrational strength of the 2-phonon state.  
Such a result has been supported by recent theoretical 
work~\cite{Ponomarev99}, where calculations showed a strong fragmentation of 
the $6^+$ member of the quadruplet.
  
The effect of the anharmonicities of the 2-phonon states in $^{208}$Pb on 
the barrier distribution was estimated with a {\tt CCFULL} calculation 
which included a reorientation term (see Eqs.~(4) and (5) in 
Ref.\ \onlinecite{Takagui90}), with the spectroscopic quadrupole moment for 
the $3^-_1$ state of $Q_{3^-_1}=-0.34$ eb~\cite{Spear83}.  The results 
are shown in 
Fig.~\ref{fig6}(a) for the case where the strength for the 2-phonon 
transition was $\sqrt{2} \beta_3$ (solid line) and when this strength was 
reduced by a factor $0.85$ [dot-dot-dashed line in Fig.~\ref{fig6}(a)].
The reduction factor applied to the pure harmonic octupole coupling 
strength was obtained from the results of Ref.\ \onlinecite{Vetter98}.
The barrier distribution from the anharmonic calculation is a slight 
improvement over the harmonic result [dashed line in Fig.~\ref{fig6}(a)] 
in region of $76$ MeV.
Any further increase in the degree of anharmonicity of the 2-phonon states 
(by reducing the energy of the 2-phonon state, for example)
leads to a barrier distribution closer in shape to the single-phonon result.
This effect is shown in Fig.~\ref{fig6}(b), where an anharmonic calculation
(solid line), with the energy of the 2-phonon at $4.424$ MeV and the 
corresponding reduction in the coupling strength of $0.28$ times that of the 
harmonic strength, is compared with the harmonic calculation 
(dashed line) and the single-phonon calculation (dot-dot-dashed line).
The reduction factor of $0.28$ was obtained in Ref.\ \onlinecite{Vetter98} 
from experimental observed intensity limits, which were then used to set 
limits relative to the expected harmonic E3 strength as a function of the 
energy of various $6^+$ states in $^{208}$Pb.

\subsection{The effects of a smaller diffuseness parameter}
\label{smalla}

As discussed earlier, the effects on the $^{16}$O+$^{208}$Pb barrier 
distribution of using a smaller diffuseness for the nuclear potential lead 
to a reduction in the height of the main barrier (an increase in its 
FWHM).  Such an effect can be explained with reference to Eq. (8) in 
Ref.\ \onlinecite{Rowley91}, since 
$d^2(E_{\mbox{\scriptsize c.m.}}\sigma)/ dE^2_{\mbox{\scriptsize c.m.}}$ 
is proportional to $\pi R_B^2/\hbar \omega _0$ (the FWHM of the main 
barrier is proportional to $\hbar \omega _0$).  
In the $^{16}$O+$^{208}$Pb reaction, a decrease in the diffuseness from 
$a=1.005$ fm to $a=0.65$ fm (resulting in an increase of $\hbar \omega _0$ 
from $3.85$ MeV to  $4.93$ MeV) led to a reduction in the height of the
main peak in the barrier distribution, as shown in Fig.~\ref{fig3}(b).
Even with this reduction to $a=0.65$, close to the value of $a$ obtained 
from fits to elastic scattering data~\cite{Thompson85}, the height of the 
main peak in the experimental barrier distribution could not be 
successfully reproduced.  

To obtain a reasonable reproduction of the measured barrier distribution, 
the diffuseness parameter had to be reduced to a value of 
$a\approx 0.40$~fm.  However, this was done at the expense of the fit to 
the high-energy fusion cross-sections (see the discussion below).  
A {\tt CCFULL} calculation with the potential parameters $V_0=283.6$ MeV, 
$r_0=1.172$ fm and $a=0.40$ fm, chosen to give an average barrier of 
$B_0=77.6$ MeV, is shown in Fig.~\ref{fig7}(b) by the dotted line.  Here 
coupling to the 2-phonon states was included with the anharmonic values of 
$4.424$ MeV for the energy of the 2-phonon states, and a reduction factor 
of $0.28$ for the 2-phonon coupling strength, as discussed earlier.  No 
transfer couplings were included in these calculations.  
After inclusion of the adiabatic $3^-_1$ state in $^{16}$O, the barrier 
distribution shown by the solid line in Fig.~\ref{fig7}(b) was obtained.  
The inclusion of the $3^-_1$ state in $^{16}$O shifts the barrier 
distribution down in energy to provide a reasonable representation of 
the data.  The third barrier distribution shown in Fig.~\ref{fig7}(b) 
(dashed line) is a CC calculation with the $a=0.40$ fm potential parameters, 
which give an average barrier of $B_0=77.6$ MeV, but {\em without} coupling 
to the 2-phonon excitations in $^{208}$Pb.  The difference between the 
2-phonon (solid line) and single-phonon (dashed line) calculations for 
$a=0.40$ fm is not as significant as the difference between the equivalent 
calculations with $a=1.005$ fm, due to the additional smoothing of the 
barrier distributions that results from the smaller diffuseness 
(larger $\hbar \omega_0$).

Such a small value for the nuclear diffuseness is problematic in that 
the experimental fusion cross-sections could not be reproduced either at 
energies above or below the average barrier.  
A diffuseness of $a=0.40$ fm, causes $\sigma$ to fall less rapidly than 
the data in the low energy region, as shown by the solid line in 
Fig.~\ref{fig7}(a).  And, in the high energy region, the calculation with 
$a=0.40$ fm significantly overestimates the data, see the inset of 
Fig.~\ref{fig7}(a).  With any of the above coupling schemes, no single 
set of potential parameters was found that could simultaneously reproduce 
the shape of the experimental barrier distribution and the fusion 
cross-sections in the low and high energy region. 

The results from the detailed CC analysis presented in this work are 
puzzling in view of the success obtained from other recent analyses of 
fusion barrier distributions~\cite{Leigh95,Dasgupta98}.  In these results, 
the shapes of the theoretical barrier distributions matched well with the 
experimental ones after including the significant couplings expected to 
affect fusion.  In contrast to this success, even after consideration of 
transfer and 2-phonon couplings in the $^{16}$O + $^{208}$Pb reaction, the 
theory was unable to reproduce the shape of the measured barrier 
distribution. 

\section{Summary and Conclusion}
\label{concl}

In this work, fission cross-sections for the $^{16}$O + $^{208}$Pb reaction 
were re-measured with improved accuracy.  The new data were found to be 
generally in good agreement with the earlier data, although some 
erroneous points in the original fission excitation function were 
identified.  The barrier distribution resulting from the new data was found 
to be a smoothly falling function for energies above the average barrier.

In order to describe the shape of the measured barrier distribution, 
detailed CC calculations were performed, avoiding where possible less 
accurate approximations often used in simplified CC analyses, and exploiting 
existing knowledge of the particle transfers in the $^{16}$O + 
$^{208}$Pb system.  It was found that coupling to the single-neutron pickup, 
single-proton and $\alpha$-stripping transfers had a significant affect on 
the barrier distribution, although coupling to these transfers in addition 
to the $3^-_1$ and $5^-_1$ single-phonon states in $^{208}$Pb, was not 
sufficient to explain the data.
Transfer from excited states in $^{208}$Pb were not included in the present 
calculations, and their effect on the shape of the barrier distribution is 
not known.  

The effects of additional coupling to 2-phonon states in $^{208}$Pb was 
explored, both in the harmonic limit and for cases that considered the 
anharmonicity of the 2-phonon states.
Inclusion of the 2-phonon states in $^{208}$Pb resulted in some improvement 
but still fell short of a complete description of the experimental barrier 
distribution.  

A better reproduction of the experimental barrier distribution was obtained 
with a very large reduction in the nuclear diffuseness parameter, from a 
value of $a=1.005$ to $a=0.40$~fm.
This approach to fitting the data was found to be unsatisfactory, since it 
destroyed the fits to the fusion cross-sections in the high and low energy 
regions.
Also, a value of $0.40$ fm for the nuclear diffuseness is significantly 
smaller than results obtained from analyses of elastic scattering 
data for the $^{16}$O + $^{208}$Pb system~\cite{Thompson89,Ball75}.  

The results from fits to the high-energy fusion cross-sections for the 
$^{16}$O + $^{208}$Pb reaction, and other systems recently 
measured~\cite{Leigh95}, also required a nuclear diffuseness larger than 
the value obtained from elastic scattering analyses.
This result indicates that the procedure for determining the potential 
parameters used in this and the work of Ref.\ \onlinecite{Leigh95} may not 
be appropriate in the analysis of fusion.  In elastic 
scattering, the more peripheral nature of the interaction means the system 
probes mainly the exponential tail of the nuclear potential.  
In contrast, fusion probes the potential at distances much closer to 
the fusion barrier radius.  In this region, the Woods-Saxon parameterisation 
may not be an adequate representation of the true nuclear potential.  
Further work is required to determine the diffuseness of the nuclear 
potential appropriate to the analysis of precise fusion data.  

Using the best available model for the description of heavy-ion fusion, it 
has been shown that the measured barrier distribution for 
$^{16}$O + $^{208}$Pb could not be reproduced with couplings to the lowest 
lying single- and 2-phonon states in $^{208}$Pb and the major particle 
transfers.  In view of the precision of the data, and the quality of the 
coupled-channels model used in its description, the disagreement between 
experiment and theory is very significant.  Further work on the appropriate
choice of the nuclear diffuseness, and a global analysis of all available 
reaction data, are required in order to improve the coupled-channels 
description of fusion for the $^{16}$O + $^{208}$Pb system.

\section*{Acknowledgements}

K.H. and I.J.T. would like to thank the Australian National University for 
their warm hospitality and partial support where this work was carried out.
M.D. acknowledges the support of a QEII Fellowship.
K.H. acknowledges the support from the Japan Society for the Promotion of
Science for Young Scientists.

\begin{table}
\begin{center}
\caption
{\small The fusion cross-sections for the \protect$^{16}$O+$^{208}$Pb 
reaction at the center-of-mass energy 
\protect$E_{\mbox{\scriptsize c.m.}}$.  }
\begin{tabular}{ccc}
E$_{\mbox{\scriptsize c.m.}}$ (MeV) & $\sigma$ (mb)  & $\delta \sigma$ (mb) 
\\ \hline
 \vspace{-0.3cm}     &     &          \\
     69.97  &   0.24  &   0.01      \\
     70.53  &   0.70  &   0.004     \\
     71.09  &   1.83  &   0.01      \\
     71.64  &   4.28  &   0.02      \\
     72.20  &   8.27  &   0.04     \\
     72.76  &   14.5  &   0.07      \\
     73.31  &   23.4  &   0.1      \\
     73.87  &   35.4  &   0.2       \\
     74.43  &   50.0  &   0.3       \\
     74.99  &  67.0   &   0.3       \\
     75.54  &  87.0   &   0.4       \\
     76.10  &  107    &   0.5       \\
     76.66  &  129    &   0.7       \\
     77.21  &  152    &   0.8       \\
     77.77  &  175    &   0.9       \\
     78.33  &  197    &   1       \\
     78.88  &  223    &   1        \\
     79.44  &  245    &   1       \\
     80.00  &  270    &   1       \\
     80.56  &  295    &   2       \\
     81.11  &  318    &   2       \\
     81.67  &  343    &   2       \\
     82.78  &  385    &   2       \\
     85.01  &  487    &   3        \\
     87.24  &  568    &   3        \\
     89.73  &  662    &   3        \\
     91.70  &  715    &   4        \\
     96.15  &  847    &   4        \\
    100.72  &  949    &   5        \\
    105.06  &  1065   &   6        \\
    109.52  &  1133   &   6      \\ \hline
\end{tabular}
\protect\label{Table1}
\end{center}
\end{table}

\begin{table}
\begin{center}
\caption
{\small The transition strengths \protect$B(E\lambda)\! \uparrow$ and 
deformation parameters \protect$\beta_{\lambda}$ for 
\protect$^{16}$O+$^{208}$Pb.  The deformation parameters were calculated with 
a nuclear radius parameter of \protect$1.06$~fm.  
The parameters for the real nuclear potential are also given. 
In the CC calculations, the nuclear deformation parameters were set to be 
equal to the Coulomb deformation parameters. }
\begin{tabular}{lccccc}
Nucleus  & $\lambda^{\pi}$ & $E^{\star}$ (MeV)  & $B(E\lambda)\! \uparrow$ & 
$\beta_{\lambda}$ & Ref. \\ \hline
 \vspace{-0.3cm}      &   &  &  &     &          \\
$^{208}$Pb &  $3^{-}_1$   &  2.615  &  
0.611 e$^{2}$b$^{3}$    &  0.161  & \cite{Spear89}  \\ 
    &  $5^{-}_1$   &  3.198  &  $\delta=0.35^{\dagger}$ fm   &  0.056 &
\cite{npa87}  \\ 
  $^{16}$O & $3^{-}_1$ & 6.129 & 0.0015 e$^{2}$b$^{3}$ & $0.733^{\ddagger}$ & 
\cite{Spear89}  \\  
 \vspace{-0.35cm}        &   &  &  &     &          \\
   &  &  $V_0$ (MeV) &  $r_0$ (fm)  &  $a$ (fm) & \\
   &  & 200.0 & 0.978 & 1.005 & \\ \hline
\end{tabular}
{\footnotesize $^{\dagger}$ here $\delta$ is the deformation length.}
{\footnotesize $^{\ddagger}$ here a nuclear radius parameter of $1.2$ fm was 
used.}
\protect\label{Table2}
\end{center}
\end{table}

\begin{figure}
\caption{ The (a) fission/ER and (b) fusion excitation functions 
for \protect$^{16}$O + $^{208}$Pb from this re-measurement 
(solid circles) and the previous data of Ref.\ \protect\onlinecite{Morton95} 
(open circles).  The ER cross-sections (open squares) are also from 
Ref.\ \protect\onlinecite{Morton95}.  }
\label{fig1}
\end{figure}

\begin{figure}
\caption{The fusion barrier distribution from this measurement (solid circles) 
compared to the previous measurement~\protect\cite{Morton95} (open symbols). 
The uncertainties associated with the barrier distribution were obtained from 
the uncertainties in the fusion cross-sections, as described in 
Ref.\ \protect\onlinecite{Leigh95}.  
See the text for an explanation of the encircled data points.  }
\label{fig2}
\end{figure}

\begin{figure}
\caption{The (a) fusion excitation functions and (b) barrier distributions 
for a single--barrier (SB) calculation (dot-dot-dashed line), and 
calculations using a single-phonon coupling scheme with two different sets of 
potential parameters [see Eq.~(1)]: \protect$V_0=200$ MeV, 
\protect$r_0=0.978$ fm, \protect$a=1.005$ fm (solid line) and 
\protect$V_0=277.5$ MeV, \protect$r_0=1.10$ fm, \protect$a=0.65$ fm 
(dashed line).  These calculations were performed with the CC code 
\protect{\tt CCFULL}. }
\label{fig3}
\end{figure}

\begin{figure}
\caption{(a) The fusion excitation function calculated using 
\protect{\tt CCFULL} \protect\cite{Hagino99} with coupling to the 
\protect$3^-_1$ and \protect$5^-_1$ single-phonon states in 
\protect$^{208}$Pb (solid line).  The dot-dot-dashed line is the same 
calculation but with the code \protect{\tt FRESCO} and with transfer included 
in addition to the single-phonon states.  
(b) The fusion barrier distribution calculated 
using \protect{\tt CCFULL} with coupling to the \protect$3^-_1$ 
and \protect$5^-_1$ single-phonon states in \protect$^{208}$Pb (solid line).  
The \protect{\tt FRESCO} result, performed with identical couplings 
(no transfer coupling), is given by the dashed line.  When transfer, in 
addition to the single-phonon states in \protect$^{208}$Pb, is included this 
results in the barrier distribution represented by the dot-dot-dashed line. }
\label{fig4}
\end{figure}

\begin{figure}
\caption{The barrier distribution for the single-phonon coupling scheme in 
\protect$^{208}$Pb (dotted line), with 2-phonon coupling (solid line), with 
single-phonon and transfer couplings (dot-dot-dashed line), and 2-phonon and 
transfer couplings (dashed line). }
\label{fig5}
\end{figure}

\begin{figure}
\caption{\small (a) The effect on the barrier distribution when the 
anharmonicity of the 2-phonon states in \protect$^{208}$Pb are taken into 
account.  The dashed line is the harmonic result where the energy of 
the 2-phonon state was 
taken as \protect$5.23$ MeV and the strength was \protect$\sqrt{2} \beta_3$. 
The barrier distribution represented by the solid line includes the 
reorientation effect with a strength unchanged from the harmonic calculation.  
The dot-dot-dashed line is the same calculation as the solid line, but the 
strength has been reduced by a factor \protect$0.85$.  Coupling to the 
transfer channels has not been included in these calculations. 
(b) The solid line is another anharmonic calculation, but assuming a lower 
energy for the 2-phonon states and with a significant reduction in the 
2-phonon coupling strength (see text).  This last result is compared with the 
same harmonic calculation shown in (a) [dashed line] and the single-phonon 
calculation (dot-dot-dashed line).  }
\label{fig6}
\end{figure}

\begin{figure}
\caption{\small (a) The fusion excitation function for a calculation with 
\protect$a=0.40$~fm and coupling to \protect$(3_1^-,5^-_1)$ single-phonon 
states in \protect$^{208}$Pb, anharmonic coupling to the 2-phonon states in 
\protect$^{208}$Pb (the strength of the 2-phonon coupling has been reduced by 
a factor \protect$0.28$ times the harmonic value, and the energy of the 
2-phonon quadruplet is \protect$4.424$ MeV), and the \protect$3^-_1$ state in
\protect$^{16}$O.  The dot-dot-dashed line is the single--barrier 
calculation with \protect$a=1.005$ fm.  The inset compares these calculations 
with the data on a linear scale.
(b) The solid line is the barrier distribution obtained from the fusion 
calculation represented by the solid line in (a).  The dotted line is the 
equivalent calculation but without the coupling to the \protect$3^-_1$ state 
in \protect$^{16}$O.  
The barrier distribution represented by the dashed line is equivalent to the
calculation represented by the solid line but now only with coupling 
to the single-phonon states in the projectile and target.  Coupling to the 
transfer channels has not been included in these calculations. }
\label{fig7}
\end{figure}

\end{document}